# Transforming magnets

Fei Sun [1, 2] and Sailing He [1, 2*]

1 Centre for Optical and Electromagnetic Research, Zhejiang Provincial Key Laboratory for Sensing Technologies, JORCEP, East Building #5,Zijingang Campus, Zhejiang University, Hangzhou 310058, China

2 Department of Electromagnetic Engineering, School of Electrical Engineering, Royal Institute of Technology (KTH), S-100 44 Stockholm, Sweden

* Corresponding author: sailing@kth.se

**Abstract**

Based on the form-invariant of Maxwell's equations under coordinate transformations, we extend the theory of transformation optics to transformation magneto-statics, which can design magnets through coordinate transformations. Some novel DC magnetic field illusions created by magnets (e.g. shirking magnets, cancelling magnets and overlapping magnets) are designed and verified by numerical simulations. Our research will open a new door to designing magnets and controlling DC magnetic fields.

## 1. Introduction

Transformation optics (TO), which has been utilized to control the path of electromagnetic waves [1-6], the conduction of current [7, 8], and the distribution of DC electric or magnetic field [9-18] in an unprecedented way, have become a very popular research topic in recent years. Based on the form-invariant of Maxwell's equation under coordinate transformations, special media (known as transformed media) with pre-designed functionality have been designed by using coordinate transformations [1-4]. By analogy to Maxwell's equations, the form-invariant of governing equations of other fields (e.g. the acoustic field and the thermal field) have been studied under coordinate transformations, and many novel devices have been designed that can control the acoustic wave [19, 20] or the thermal field [21, 22].

The current theory of TO can be directly utilized to design some passive magnetic media (with transformed permeability) to control the distribution of the DC magnetic field: a DC magnetic cloak that can hide any objects from being detected from the external DC magnetic field [16, 17], a DC magnetic concentrator that can achieve an enhanced DC magnetic field with high uniformity [9-13], a DC magnetic lens that can both amplify the background DC magnetic field and the gradient of the field [14, 15], and a carpet DC magnetic field [18]. However there has been no study on transforming magnets by using TO, and no literature describes how the residual induction (or intensity of magnetization) of a magnet transforms if there is a magnet in the reference space.

In this paper we extend the current TO to the case involving the transforming of the residual induction of a magnet. Based on the proposed theory, we design three novel devices that can create the illusion of magnets: the first one shrinks the magnet (e.g. a small magnet can perform like a bigger one); the second nullifies the magnet (e.g. we can cancel the DC magnetic field produced by a magnet by adding special anti-magnets and transformed materials); and the third overlaps magnets (e.g. we can overlap magnets in different spatial positions to perform effectively like one magnet with overlapped residual induction). The study in this paper will lead a new way to design magnets and create the illusion of a DC magnetic field.

**2. Theory method**

In the section, we will extend the theory of TO to transform the magnetization intensity of magnets. Our starting point is two magnetic field equations in Maxwell's equations. In the reference space of a Cartesian coordinate system, we can write them as [4]:

$$\begin{cases} \left(B^i\right)_{,i} = 0 \\ [ijk]H_{k,j} = J^i \end{cases}. \quad (1)$$

The comma refers to partial differentiation. [ijk] is the permutation symbol, which is the same as the Levi-Civita tensor in the right-hand Cartesian coordinate system. Due to the form-invariant of Maxwell's equations (any physics equation expressed in tensor form is form-invariant under coordinate transformations), we can rewrite Eq. (1) in a curved coordinate space as [4]:

$$\begin{cases} \left(\sqrt{g'}B^{i'}\right)_{,i'} = 0 \\ [i'j'k']H_{k',j'} = \sqrt{g'}J^{i'} \end{cases}. \quad (2)$$

where $g'=\det(g_{i'j'})$. $g_{i'j'}$ is metric tensor in this curved coordinate system. The key point of TO is material interpretation [1, 4]: we can assume that the electromagnetic medium and the curved coordinates have an equivalent effect on the electromagnetic wave, and treat Eq. (2) as Maxwell's equations in a Cartesian coordinate system of a flat space but not in a curved coordinate space. We make the following definitions:

$$\begin{cases} \tilde{B}^{i'} = \sqrt{g'}B^{i'} = \sqrt{g'}\Lambda^{i'}{}_i B^i \\ \tilde{H}_{i'} = H_{i'} = \Lambda_{i'}{}^i H_i \\ \tilde{J}^{i'} = \sqrt{g'}J^{i'} = \sqrt{g'}\Lambda^{i'}{}_i J^i \end{cases}. \quad (3)$$

The quantities with superscript tilde '~' and primes indicate that the quantities are in the real/physical space; the quantities without primes indicate the ones in the reference space (a virtual space), and the quantities with primes and without the superscript '~' are ones in the transition space (a curved space). In the real space, the space is flat while filled with some special medium (the transformed medium), which will be deduced later. In order to express quantities conveniently, we often drop the superscript '~' in the real space, and rewrite Eq. (3) as:

$$\begin{cases} B^{i'} = \sqrt{g'}\Lambda^{i'}{}_i B^i \\ H_{i'} = \Lambda_{i'}{}^i H_i \\ J^{i'} = \sqrt{g'}\Lambda^{i'}{}_i J^i \end{cases}. \quad (4)$$

Here the quantities with or without primes indicate the ones in the real or reference space, respectively. Since the reference space is in the Cartesian coordinate system (which means $g=1$), the metric tensor in Eq. (4) can be rewritten as:

$$\sqrt{g'} = \sqrt{\det(g_{i'j'})} = \sqrt{\det(\Lambda_{i'}{}^i \Lambda_{j'}{}^j g_{ij})} = \det(\Lambda_{i'}{}^i)\sqrt{g} = \det(\Lambda_{i'}{}^i) = \frac{1}{\det(\Lambda^{i'}{}_i)}. \quad (5)$$

By using Eq. (5), we can rewrite Eq. (4) as:

$$\begin{cases} B^{i'} = \dfrac{1}{\det(\Lambda^{i'}{}_i)} \Lambda^{i'}{}_i B^i \\ H_{i'} = \Lambda_{i'}{}^i H_i \\ J^{i'} = \dfrac{1}{\det(\Lambda^{i'}{}_i)} \Lambda^{i'}{}_i J^i \end{cases} \quad (6)$$

Note Eq. (6) can also be deduced by other ways (e.g. multivariable calculus [6]). In traditional TO, we often assume that the medium in the reference space does not contain any magnets and this means the residual induction $\boldsymbol{B}_r$ (or intensity of magnetization $\boldsymbol{M}=\boldsymbol{B}_r/\mu_0$) is zero everywhere. In this case, we have:

$$\begin{cases} B^{i'} = \mu_0 \mu^{i'j'} H_{j'} \\ B^i = \mu_0 \mu^{ij} H_j \end{cases} \quad (7)$$

Combining Eq. (6) and (7), we can obtain the relationship between the medium in the real space and the reference space:

$$\mu^{i'j'} = \dfrac{1}{\det(\Lambda^{i'}{}_i)} \Lambda^{i'}{}_i \Lambda^{j'}{}_j \mu^{ij}. \quad (8)$$

Eq. (8) is a classical equation in TO. If there are some magnets in the reference space, then Eq. (7) should be modified as:

$$\begin{cases} B^{i'} = \mu_0 \mu^{i'j'} H_{j'} + \mu_0 M^{i'} \\ B^i = \mu_0 \mu^{ij} H_j + \mu_0 M^i \end{cases} \quad (9)$$

Here $\boldsymbol{M}$ and $\boldsymbol{M'}$ correspond to the magnetization intensity of magnets in the reference and real space, respectively. In this case, we can combine Eq. (6) and (9) to obtain the following relation:

$$\dfrac{1}{\det(\Lambda^{i'}{}_i)} \Lambda^{i'}{}_i \Lambda^{j'}{}_j \mu^{ij} H_{j'} + \dfrac{1}{\det(\Lambda^{i'}{}_i)} \Lambda^{i'}{}_i M^i \equiv \mu^{i'j'} H_{j'} + M^{i'}. \quad (10)$$

Considering that Eq. (10) is true for any magnetic field $H'$, we can obtain:

$$\begin{cases} \mu^{i'j'} = \dfrac{1}{\det(\Lambda^{i'}{}_i)} \Lambda^{i'}{}_i \Lambda^{j'}{}_j \mu^{ij} \\ M^{i'} = \dfrac{1}{\det(\Lambda^{i'}{}_i)} \Lambda^{i'}{}_i M^i \end{cases} \quad (11)$$

Eq. (11) gives the complete transformation of the relation between the magnetic materials in the reference space and the transformed magnetic materials in the real space, even if there are some magnets in the reference space. As we can see from Eq. (11), if there is no magnet in the reference space $\boldsymbol{M}=0$, we have $\boldsymbol{M'}=0$. In this case, only permeability needs to be transformed (Eq. (11) reduces to Eq. (8)), which is consistent with classical TO. We can rewrite Eq. (11) in a matrix form:

$$\begin{cases} \overline{\overline{\mu}}' = \dfrac{1}{\det(\Lambda)} \Lambda \overline{\overline{\mu}} \Lambda^T \\ \overline{M}' = \dfrac{1}{\det(\Lambda)} \Lambda \overline{M} \end{cases}, \quad (12)$$

where $\Lambda = \partial(x',y',z')/\partial(x,y,z)$ is the Jacobian transformation matrix. Now we have extended the classical TO to the case in which there are magnets in the reference space. In the next section, we will use this theory to design some novel devices which can create illusions of magnets.

**3. Examples**

1) Rescaling a magnet.

The first example is a device that can amplify the volume of a magnet. Fig. 1 (a) shows the basic idea of this illusion. For simplicity, we consider a 2D circular magnet with radius $R_2$ filled in the free space (infinitely long in the $z$-direction; see the left part of Fig. 1(b)). The relative permeability and magnetization intensity in the reference space can be given, respectively, as:

$$\mu = \begin{cases} \mu_{m0}, & r \in [0, R_2] \\ 1, & r \in [R_2, R_3]; \\ 1, & r \in [R_3, \infty) \end{cases} \quad (13)$$

and

$$\overline{M} = \begin{cases} \dfrac{\overline{B}_{r0}}{\mu_0}, & r \in [0, R_2] \\ 0, & r \in [R_2, R_3]. \\ 0, & r \in [R_3, \infty) \end{cases} \quad (14)$$

$\boldsymbol{B}_{r0}$ is the residual induction in the reference space. $\mu_{m0}$ is the relative permeability of the magnet in the reference space. Now we want to use a magnet (of the same $\mu_{m0}$) with a smaller radius $R_1$ ($R_1<R_2$) in the real space to produce the same DC magnetic field in the region $r'>R_3$ (as if it is produced by a big magnet with radius $R_2$ in the reference space). The reference space is shown in the left part of Fig. 1(b). The permeability and magnetization distributions are given in Eq. (13) and (14). The real space is shown in the right part of Fig. 1(b). The magnet in the region $0<r<R_2$ in the reference space is compressed to the region $0<r'<R_1<R_2$ in the real space. The air region $R_2<r<R_3$ in the reference space is expanded to the region $R_1<r<R_3$ in the real space while keeping the air region $r>R_3$ in the reference space identical to the region $r'>R_3$ in the real space. The full transformation can be given as:

$$r' = \begin{cases} R_1 r / R_2 & ,r \in [0, R_2] \\ (R_3 - R_1)r/(R_3 - R_2) + R_3(R_1 - R_2)/(R_3 - R_2), & r \in [R_2, R_3]; \theta' = \theta; z' = z. \\ r & ,r \in [R_2, \infty) \end{cases} \quad (15)$$

Combing Eq. (11) and (15), we can obtain the relative permeability and magnetization intensity

in the real space:

$$\overline{M}' = \begin{cases} diag(\frac{R_2}{R_1}, \frac{R_2}{R_1}, \left(\frac{R_2}{R_1}\right)^2) \frac{\overline{B}_{r0}}{\mu_0}, & r' \in [0, R_1] \\ 0, & r' \in [R_1, R_3]; \\ 0, & r' \in [R_3, \infty) \end{cases} \quad (16)$$

$$\overline{\overline{\mu}}' = \begin{cases} \mu_{m0} diag(1,1,(R_2/R_1)^2) & ,r' \in [0, R_1] \\ diag(\frac{(R_3-R_2)r'+(R_2-R_1)R_3}{(R_3-R_2)r'}, \frac{(R_3-R_2)r'}{(R_3-R_2)r'+(R_2-R_1)R_3}, \frac{(R_3-R_2)[(R_3-R_2)r'+(R_2-R_1)R_3]}{(R_3-R_1)^2 r'}), & ,r' \in [R_1, R_3]. \\ 1 & ,r' \in [R_3, \infty) \end{cases} \quad (17)$$

The above quantities are expressed in the cylindrical coordinate system. For the 2D case (the magnetic field is in the plane $z=0$ and $\boldsymbol{B}_{r0}$ does not contain any z-component), we can just drop the z-component in the above expression, and they will reduce to:

$$\overline{M}' = \begin{cases} \frac{R_2}{R_1} \frac{\overline{B}_{r0}}{\mu_0}, & r' \in [0, R_1] \\ 0, & r' \in [R_1, R_3] \\ 0, & r' \in [R_3, \infty) \end{cases} \quad (18)$$

and

$$\overline{\overline{\mu}}' = \begin{cases} \mu_{m0} & ,r' \in [0, R_1] \\ diag(\frac{(R_3-R_2)r'+(R_2-R_1)R_3}{(R_3-R_2)r'}, \frac{(R_3-R_2)r'}{(R_3-R_2)r'+(R_2-R_1)R_3}), & ,r' \in [R_1, R_3]. \\ 1 & ,r' \in [R_3, \infty) \end{cases} \quad (19)$$

Note that the quantities in Eq. (18) and (19) are expressed in a 2D cylindrical coordinate system. As we can see from Eq. (18), if we want a magnet with small radius $R_1$ to produce the same DC magnetic field distribution as a magnet (of the same relative permeability) with a larger radius $R_2$, its magnetization intensity should be rescaled accordingly. We use the finite element method (FEM) to verify the performance of the device (see Fig. 1(c) and (d)). The distributions of the DC magnetic field outside the region $r'>R_3$ are exactly the same for the case where we only have a magnet with big radius $R_2$ (see Fig. 1(c); permeability and magnetization distributions are given in Eq. (13) and (14), respectively) and the case where we have a magnet (of same relative permeability) with small radius $R_1$ and some rescaling medium (see Fig. 1(d); permeability and magnetization distributions are given in Eq. (18) and (19), respectively). The FEM simulation is performed by using a commercial software, COMSOL Multiphysics. The distribution of the relative permeability of the scaling medium ($R_1<r'<R_3$) in Eq. (19) is shown in Fig. 1(e) and (f): the permeability along the radial and tangential directions are both larger than zero, which can be realized by using two isotropic materials (one with high permeability and the other with low permeability) layer by layer along the radial direction (similar to the method to creating a magnetic concentrator for DC magnetic field enhancement [9, 13]).

2) Cancelling magnets.

Analogous to the scattering-cancelling by using a complementary medium in the electromagnetic wave case [3], we can design complementary magnets and complementary media to cancel the DC magnetic field produced by some magnets (see Fig. 2(a) and (b)). We design a specific structure shown in Fig. 2(c) with $h = d = 0.2$ m. The coordinate transformation is given as the following.

The yellow and red regions:

$$x' = -\frac{x}{2}; \quad y' = y; \quad z' = z; \quad (20\text{-}1)$$

The purple region:

$$\begin{cases} x' = [u(y+d) - u(y-d)](x/4 + 3d/4) + u(y-d)[-3y/4 + x/4 + 3d/2] + u(-y-d)[3y/4 + x/4 + 3d/2] \\ y' = y \\ z' = z \end{cases} \quad (20\text{-}2)$$

And the other white region and green region:

$$x' = x; \quad y' = y; \quad z' = z; \quad (20\text{-}3)$$

where function $u$ is defined as follows:

$$u(\xi) = \begin{cases} 1, \xi \geq 0 \\ 0, \xi < 0 \end{cases} \quad (21)$$

Combining Eq. (20) and Eq. (11), we can obtain the material in:
the yellow region (complementary medium):

$$\mu'_{yellow} = diag(-\frac{1}{2}, -2, -2); \quad (22)$$

the red region (the complementary magnet):

$$\begin{cases} \mu'_{red} = \mu_{green} diag(-\frac{1}{2}, -2, -2) \\ \overline{M}'_{red} = diag(1, -2, -2) \overline{M}_{green} \end{cases}, \quad (23)$$

where $\mu_{green}$ and $M_{green}$ are the relative permeability and magnetization intensity of the green magnet, respectively;
and the purple region (restoring medium):

$$\begin{cases} \mu'_{purple} = \begin{bmatrix} 2.5 & 3 & 0 \\ 3 & 4 & 0 \\ 0 & 0 & 4 \end{bmatrix}, y \in [-2d, -d], x' \in [0.5d, d] \\ \mu'_{purple} = diag(0.25, 4, 4), y \in [-d, d], x' \in [0.5d, d]. \\ \mu'_{purple} = \begin{bmatrix} 2.5 & -3 & 0 \\ -3 & 4 & 0 \\ 0 & 0 & 4 \end{bmatrix}, y \in [d, 2d], x' \in [0.5d, d] \end{cases} \quad (24)$$

The FEM simulation results are shown in Fig. 2(d) and (e): if we only put one magnet (the green part in Fig, 2(c)), the DC magnetic field is non-zero outside the white region in Fig. 2(d). However if we add some complementary magnet (red region), complementary medium (yellow region), and restoring medium (purple region) to the original magnet (green region), the DC magnetic field can be greatly reduced outside the whole structure (see Fig. 2(e)).

We note that people can shield the DC magnetic field around a magnet simply by adding a magnetic insulation layer (e.g. a superconductor shell). In this case, the region without the magnetic field and the region with the magnetic field are completely isolated. Our method is different: we do not need to form a closed region by using a magnetic insulation shell, but instead we put a complementary magnet and complementary medium aside the magnet. Therefore, the magnet region is connected with the outside region (e.g. other objects can be moved freely between the region without the magnetic field and the region with the magnetic field).

3) Overlapped magnets

Inspired by overlapped illusions of light sources [5], we can also create an illusion of overlapped magnets. The basic idea is shown in Fig. 3 (a): two magnets at different locations can perform like one magnet with overlapped magnetization intensity to the outside observer. To create such an illusion, we need to transform both the permeability and magnetization intensity, which require the extended TO in this paper (Eq. (11)).

The structure we chose here is similar to the second example for cancelling magnets (see Fig. 3(b)): we use the same coordinate transformation Eq. (20) for each region (the only difference is that we remove the complementary magnet in the red region in Fig. 2(b) and add another magnet in the blue region in Fig. 3(b)). The coordinate transformation in the blue region (filled with a magnet) is given in Eq. (20-2). We still choose $h = d = 0.2$ m., so the materials in the yellow and purple regions are still described by Eq. (22) and (24), respectively. The material in the blue region can be given as:

$$\begin{cases} \mu'_{blue} = \mu_{green} diag(0.25, 4, 4) \\ \overline{M}'_{blue} = diag(1, 4, 4) \overline{M}_{green} \end{cases}. \quad (25)$$

In this case, the blue magnet, whose permeability and magnetization can be determined by Eq. (25) together with purple and yellow media (whose permeabilities are given in Eq. (22) and (24)), will produce the same DC magnetic field as that produced by only one green magnet with permeability $\mu_{green}$ and magnetization $M_{green}$ in the external space. The simulation results are given in Fig. 3(c) to (f). As we can see from Fig. 3(c) and (d), the distributions of amplitude of DC magnetic flux are the same for the case where only one green magnet is used and the case where only one blue magnet (the residual inductions are both 10 T along the *x*-direction) with transformed media in the yellow and purple regions introduced. Fig. 3(e) and (f) show that if we introduce both the green magnet and blue magnet with a residual induction of 10 T along the x-direction, and a transformed medium in the yellow and purple regions, the distribution of amplitude of DC magnetic flux outside the whole structure is the same as when we only introduce a green magnet with a residual induction of 20 T along the *x*-direction.

**Summary**

We extend the theory of transformation optics to include the transformation of magnets. The proposed theory will usher in a new way to design magnets: a coordinate transformation method. By using coordinate transformations, we can create many illusions of magnets, such as rescaling magnets, cancelling magnets, and overlapping magnets. These novel devices, which can manipulate the DC magnetic field produced by magnets, will have many potential applications (e.g. magnetic field shielding, magnetic sensors, etc.).

**Acknowledgement**

This work is partially supported by the National High Technology Research and Development Program (863 Program) of China (No. 2012AA030402), the National Natural Science Foundation of China (Nos. 61178062 and 60990322), the Program of Zhejiang Leading Team of Science and Technology Innovation, Swedish VR grant (# 621-2011-4620) and SOARD. Fei Sun thanks the China Scholarship Council (CSC) NO. 201206320083.

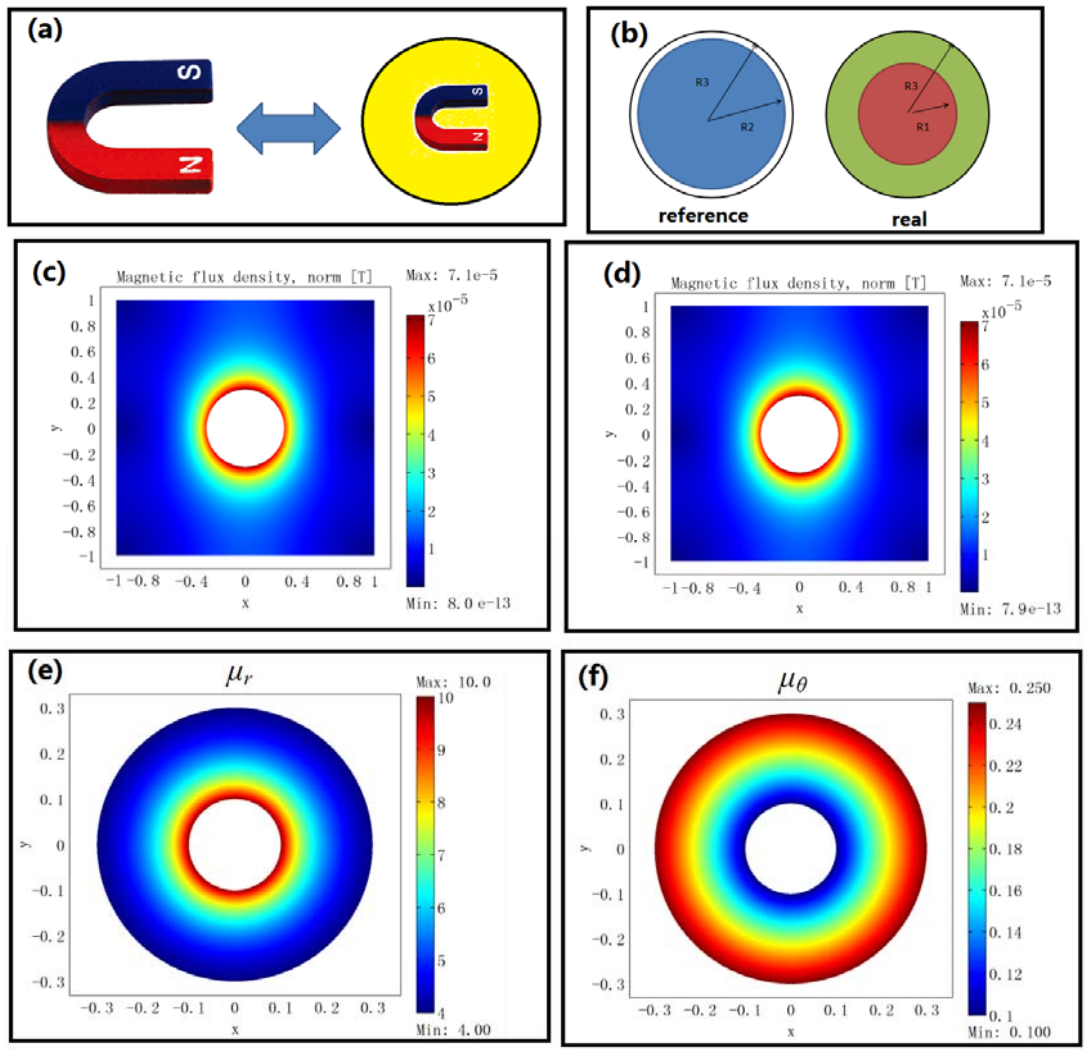

Fig. 1. (a) The basic idea of this rescaling magnet illusion is that we can use a small magnet and a transformed medium to cause the whole structure to perform like a big magnet. (b) The transformation relation for a 2D cylindrical magnet. The left part is the reference space: the blue circle is the big magnet with radius $R_2$ filled in air (white region). The right part is the real space: the red circle is the small magnet with radius $R_1$, the green region is the rescaling medium ($R_1<r'<R_3$ given in Eq. (19)), and the remaining white region is air. (c) FEM simulation result: the magnetic flux distribution in the region $r'>R_3 = 0.3$ m produced by a single magnet with radius $R_2 = 0.25$ m, relative permeability $\mu_{m0} = 1000$ and a residual induction of 0.1 T along the $x$-direction. (d) FEM simulation result: the magnetic flux distribution in the region $r'>R_3 = 0.3$ m produced by a magnet with radius $R_1 = 0.1$ m, same relative permeability $\mu_{m0}=1000$ and a residual induction of 0.25 T (determined by Eq. (18)) along the $x$-direction and rescaling medium described by Eq. (19) in the region $R_1<r'<R_3$. (e) The distribution of relative permeability along the radial direction for the rescaling medium in (d). (f) The distribution of relative permeability along the tangential direction for the rescaling medium in (d).

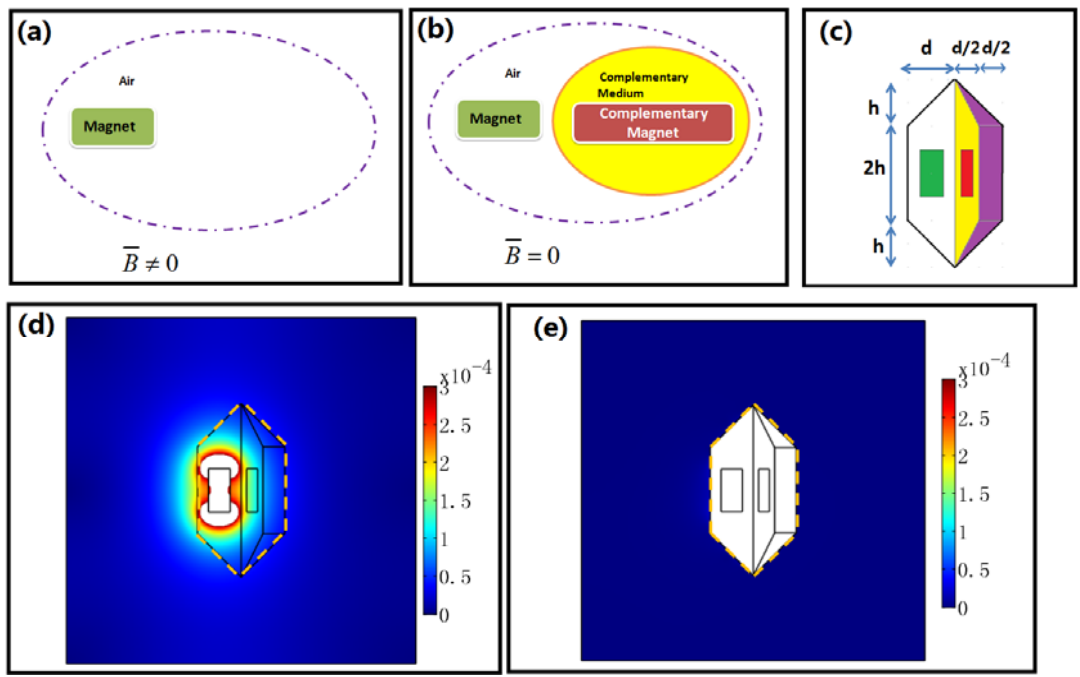

Fig. 2. (a) and (b): The basic idea of cancelling magnets by using complementary magnets and a complementary medium. (a) One magnet can produce a non-zero DC magnetic field in a certain region of space (e.g. the region outside the dashed purple line). (b) If we put a complementary magnet and complementary medium aside the original magnet, the DC magnetic field in the same region (e.g. the region outside the dashed purple line) becomes nearly zero. (c) A specific structure with $h = d = 0.2$ m to illustrate our idea of a cancelling magnet: the green region is the magnet with relative permeability 1000, residual induction 1 T along the $x$-direction, width 0.1 m and height 0.2 m; the red region is the complementary magnet with relative permeability and residual induction calculated from Eq. (23), width 0.05 m and height 0.2 m; the yellow region is the complementary medium with relative permeability described by Eq. (22); and the purple region is the restoring medium with relative permeability described by Eq. (24). (d) and (e): the FEM simulation results. (d) The amplitude of the total DC magnetic flux density distribution outside the whole structure when only one magnet (green part in (c)) is used. (e) The amplitude of the total

DC magnetic flux density distribution outside the whole structure when the original magnet (green region), complementary magnet (red region), complementary medium (yellow region) and restoring medium (purple region) are all used.

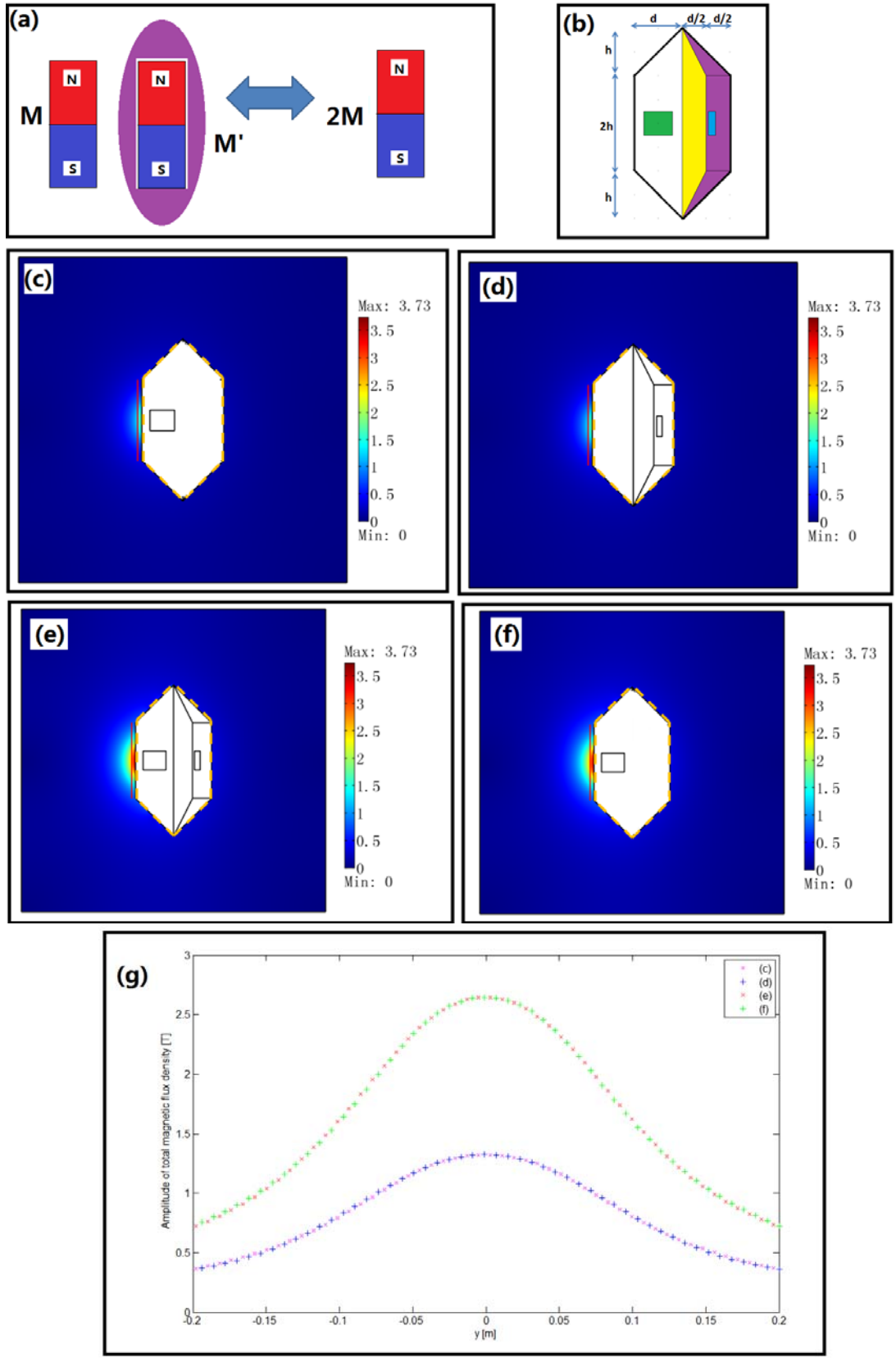

Fig. 3. (a) The basic idea of overlapped magnets: two magnets at different locations with some transformed medium can mimic one magnet of higher magnetization intensity for outside observers. (b) A specific structure with $h = d = 0.2$ m: the green region is one magnet with width

0.12 m and height 0.1 m; the blue region is another magnet with width 0.03 m and height 0.1 m; the yellow region is the complementary medium with relative permeability described by Eq. (22); the purple region is the restoring medium with relative permeability described by Eq. (24). (c) The amplitude of the total DC magnetic flux density distribution outside the whole structure when only the green magnet with relative permeability $\mu_{green}= 1$ and residual induction 10 T along the *x*-direction is used. Note that in this case we do not introduce the blue magnet and transformed media in the yellow and purple regions (they are all set as air). (d) The amplitude of the total DC magnetic flux density distribution outside the whole structure with only the blue magnet with the relative permeability and residual induction calculated from Eq. (25) by choosing $\mu_{green}=1$ and $\boldsymbol{M}_{green}=\mathrm{diag}(10\ \mathrm{T},0,0)/\mu_0$. The transformed media in the yellow and purple regions are introduced (calculated by Eq. (22) and (24)). Note that in this case we do not introduce the green magnet. (e) The amplitude of the total DC magnetic flux density distribution outside the whole structure when both green and blue magnets are introduced (the relative permeability and the residual induction of these magnets are the same as those in (c) and (d)). We also introduce transformed media in the yellow and purple regions. (f) The amplitude of the total DC magnetic flux density distribution outside the whole structure when only the green magnet with double residual induction in (c) is introduced. In this case, yellow, purple, and blue regions are set as air. (g) The amplitude of the total DC magnetic flux density distribution along the same red line indicated in (c) to (f), which also reveals that the magnetic flux density distributions in (c) and (d) are exactly the same, and the magnetic flux density distributions in (e) and (f) are exactly the same.